\documentclass[11pt]{article}

\usepackage{times}
\usepackage{amsmath, amssymb}
\usepackage{amsthm}
\usepackage{algorithm}
\usepackage{algorithmic}
\newcommand{\E}{\mathrm{E}}
\newcommand{\eps}{\epsilon}
\newcommand{\bR}{\mathbb{R}}

\usepackage{color}

\newcommand{\D}{{\cal D}}
\def\argmin{\mathop{\rm argmin}}

\newtheorem{theorem}{Theorem}[section]

\newtheorem{corollary}[theorem]{Corollary}
\newtheorem{observation}[theorem]{Observation}
\newtheorem{lemma}[theorem]{Lemma}
\newtheorem{proposition}[theorem]{Property}

\theoremstyle{remark}

\newtheorem{claim}[theorem]{Claim}

\theoremstyle{definition}
\newtheorem{definition}[theorem]{Definition}

\title{A Learning Theory Approach to \\ Non-Interactive Database Privacy}
\author{Avrim Blum\thanks{Department of Computer Science, Carnegie Mellon University. Email: {\tt avrim@cs.cmu.edu}}\and 
Katrina Ligett\thanks{Department of Computer Science, California Institute of Technology. Email: {\tt katrina@caltech.edu}}\and
Aaron Roth\thanks{Department of Computer and Information Science, University of Pennsylvania. Email: {\tt aaroth@cis.upenn.edu}}
}

\begin{document}
%\markboth{Blum, Ligett, Roth}{A Learning Theory Approach to Non-Interactive Database Privacy}

\maketitle

\begin{abstract}
In this paper we demonstrate that, ignoring computational constraints, it is possible to privately release synthetic databases that are useful for large classes of queries -- much larger in size than the database itself. Specifically, we give a mechanism that privately releases synthetic data for a class of queries over a discrete domain with error that grows as a function of the size of the smallest net approximately representing the answers to that class of queries. We show that this in particular implies a mechanism for counting queries that gives error guarantees that grow only with the VC-dimension of the class of queries, which itself grows only logarithmically with the size of the query class.

We also show that it is not possible to privately release even simple classes of queries (such as intervals and their generalizations) over continuous domains. Despite this, we give a privacy-preserving polynomial time algorithm that releases information useful for all halfspace queries, given a slight relaxation of the utility guarantee.  This algorithm does not release synthetic data, but instead another data structure capable of representing an answer for each query. We also give an efficient algorithm for releasing synthetic data for the class of interval queries and axis-aligned rectangles of constant dimension.

Finally, inspired by learning theory, we introduce a new notion of data privacy, which we call {\em{distributional privacy}}, and show that it is strictly stronger than the prevailing privacy notion, differential privacy.

\end{abstract}

%\category{F.2}{Theory of Computation}{Analysis of Algorithms and Problem Complexity}
%\terms{Algorithms, Security, Theory}
%\keywords{non-interactive database privacy, learning theory}

%\thispagestyle{empty} \setcounter{page}{0}
%\clearpage

\section{Introduction}
As large-scale collection of personal information becomes easier, the
problem of database privacy is increasingly important. In many cases, we
might hope to learn useful information from sensitive data (for example,
we might learn a correlation between smoking and lung cancer from a
collection of medical records). However, for legal, financial, or moral
reasons, administrators of sensitive datasets might not want to release
their data. If those with the expertise to learn from large datasets are
not the same as those who administer the datasets, what is to be done?
In order to study this problem theoretically, it is important to
quantify what exactly we mean by ``privacy.''

A series of recent papers \cite{DN04,BDMN05,DMNS06} formalizes the
notion of \textit{differential privacy}. A database privatization
mechanism (which may be either interactive or non-interactive)
satisfies differential privacy if the addition or removal of a single
database element does not change the probability of any outcome of the
privatization mechanism by more than some small amount. The definition
is intended to capture the notion that ``distributional information is
not private''---we may reveal that smoking correlates to lung cancer,
but not that any individual has lung cancer. Individuals may submit
their personal information to the database secure in the knowledge
that (almost) nothing can be discovered from the database with their
information that could not
have been discovered without their information.

%There has been a series of lower bound results
%\cite{DN03,DMNS06,DMT07} that suggest that non-interactive databases
%(or interactive databases that can be queried a linear number of
%times) cannot accurately answer all queries, or an adversary will be
%able to reconstruct all but a $1-o(1)$ fraction of the original
%database exactly (obviously a very strong violation of privacy). As a
%result, most recent work has focused on the design of interactive
%mechanisms that answer only a sublinear number of queries. However,
%since these mechanisms may only answer a sublinear number of queries
%\textit{in total} (not per user), after which point they must be
%destroyed, this limits their practicality
%%in some situations
%{in situations where the number of queries that might be asked is
%comparable to or larger than the number of entries in the database}.

In this paper, motivated by learning theory, we propose the study of
privacy-preserving mechanisms that are useful
for all queries in a particular class (such as all conjunction queries
or all halfspace queries).
In particular, we focus on counting queries of the form, ``what
fraction of the database entries satisfy predicate $\varphi$?'' and say
that a sanitized output is useful for a class $C$ if the answers to all
queries in $C$ have changed by at most some $\pm \alpha$.

%In doing
%so, we circumvent existing lower bounds for non-interactive databases
%that only hold for particular types of queries, such as subset sum
%queries \cite{DN03,DMNS06,DMT07}.

Building on the techniques of Kasiviswanathan et al.~\cite{KLNRS08}, we show that for discretized domains, for any concept class that admits an $\alpha$-net $\mathcal{N}_\alpha$, it is possible to privately release synthetic data that is useful for the class, with error that grows proportionally to the \emph{logarithm} of the size of $\mathcal{N}_\alpha$. As a consequence, we show that it is possible to release data useful for a set of counting queries with error that grows proportionally to the VC-dimension of the class of queries. The algorithm is not in general computationally efficient. We are able to give a different algorithm that efficiently releases synthetic data for the class of interval queries (and more generally, axis-aligned rectangles in fixed dimension) that achieves guarantees in a similar range of parameters.

Unfortunately, we show that {for non-discretized domains}, under
the above definition of usefulness, it is impossible to publish a
differentially private database that is useful for
even quite simple classes such as interval queries.
% over a non-discretized domain.
We next show how, under a natural relaxation of the usefulness
criterion, one can release information that can be used to usefully
answer (arbitrarily many) halfspace queries while satisfying
privacy. In particular, instead of requiring that useful mechanisms
answer each query approximately correctly, we allow our algorithm to
produce an answer that is approximately correct {\em{for some nearby
query}}. {This relaxation is motivated by the notion of
large-margin separators in learning theory
\cite{AB99,Vapnik:book98,SVMbook}; in particular, queries with no
data points close to the separating hyperplane must be answered
accurately, and the allowable error more generally is a function of the
fraction of points close to the hyperplane.}

We also introduce a new concept, \textit{distributional privacy}, which makes
explicit the notion that when run on a database drawn from a
distribution, privacy-preserving mechanisms should reveal only
information about the underlying distribution, and nothing else.
Given a distribution $\mathcal{D}$ over database points, a database
privatization mechanism satisfies distributional privacy if with high
probability, drawing an entirely new database from $\mathcal{D}$ does not change
the probability of any outcome of the privatization mechanism by more
than some small amount. We show that distributional privacy is a
strictly stronger guarantee than differential privacy by showing that
any mechanism that satisfies distributional privacy also satisfies
differential privacy, but that there are some functions that can be
answered accurately while satisfying differential privacy, and yet
reveal information about the particular database (although not about
any particular database element) that is not ``distributional.''

\subsection{Prior and Subsequent Work}
\subsubsection{Prior Work}
Recent work on theoretical guarantees for data privacy was initiated
by \cite{DN03}.  The notion of differential privacy, finally formalized by
\cite{DMNS06},
separates issues of privacy from issues of outside information by
defining privacy as indistinguishability of neighboring databases.
This captures the notion that (nearly) anything that can be learned if
your data is included in the database can also be learned without your
data.  This notion of privacy ensures that users have very little
incentive to withhold their information from the database.  The
connection between data privacy and incentive-compatibility was formalized by McSherry and Talwar \cite{MT07}.

%The majority of the work on non-interactive mechanisms has yielded
%impossibility results \cite{DN03,DMNS06,DMT07}.
Much of the initial work focused on \emph{lower bounds}. Dinur and
Nissim \cite{DN03} showed that any mechanism that answers substantially more than a linear number of \emph{subset-sum} queries with error $o(1/\sqrt{n})$ yields what they called \emph{blatant non-privacy} -- i.e. it allows an adversary to reconstruct all but a $o(1)$ fraction of the original database. Dwork et al.\
\cite{DMT07} extend this result to the case in which the private mechanism can answer a constant fraction of queries with arbitrary error, and show that still if the error on the remaining queries is $o(1/\sqrt{n})$, the result is blatant non-privacy. Dwork and Yekhanin \cite{DY08} give further improvements. These results easily extend to the case of counting queries which we consider here.

%As a result of these lower bounds in the non-interactive setting,
%recent work has focused on interactive mechanisms.
Dwork et al.\ \cite{DMNS06}, in the paper that defined differential privacy, show that releasing the answers to $k$ \emph{low sensitivity} queries with noise drawn independently from the Laplace distribution with scale $k/\epsilon$ preserves $\epsilon$-differential privacy. Unfortunately, the noise scales linearly in the number of queries answered, and so this mechanism can only answer a sub-linear number of queries with non-trivial accuracy.  Blum et
al.\ \cite{BDMN05} consider a model of learning and show that concept
classes that are learnable in the statistical query (SQ) model are
also learnable from a polynomially sized dataset accessed through an
interactive differential-privacy-preserving mechanism.  We note that
such mechanisms still access the database by asking counting-queries perturbed with independent noise from the Laplace distribution, and so can still only make a sublinear number of queries. In this paper, we give a mechanism for privately answering counting queries with noise that grows only logarithmically with the number of queries asked (or more generally with the VC-dimension of the query class). This improvement allows an analyst to answer an exponentially large number of queries with non-trivial error, rather than only linearly many.

Most similar to this paper is the work of Kasiviswanathan et
al.\ \cite{KLNRS08}. Kasiviswanathan et al.\ study what can be learned
privately when what is desired is that the hypothesis output by the
learning algorithm satisfies differential privacy. They show that in a
PAC learning model in which the learner has access to the private
database, ignoring computational constraints, anything that is PAC
learnable is also privately PAC learnable. We build upon the technique
in their paper to show that in fact, it is possible to privately
release a dataset that is simultaneously useful for any function in a
concept class of polynomial VC-dimension.  Kasiviswanathan et al.\ also study several restrictions on
learning algorithms, show separation between these learning models,
and give efficient algorithms for learning particular concept classes.
%
%In this work, we study non-interactive database release mechanisms,
%which may be used to privately release a fixed database or datastructure which is useful for answering very large classes of statistical queries without the need for further involvement of the mechanism.  Blum et al.\ \cite{BDMN05} consider
%running machine learning algorithms on datasets that are accessed
%through interactive privacy-preserving mechanisms.
%In contrast, we show how to release data sets from which one can
%usefully learn the values of all functions in restricted concept classes.

\subsubsection{Subsequent Work}
Since the original publication of this paper in STOC 2008 \cite{BLR08} there has been a substantial amount of follow up work. A sequence of papers by Dwork et al. \cite{DNRRV09,DRV10} give a non-interactive mechanism for releasing counting queries with accuracy that depends in a similar way to the mechanism presented in this paper on the total number of queries asked, but has a better dependence on the database size. This comes at the expense of relaxing the notion of $\epsilon$-differential privacy to an approximate version called $(\epsilon,\delta)$-differential privacy. The mechanism of \cite{DRV10} also extends to arbitrary low-sensitivity queries rather than only counting queries. This extension makes crucial use of the relaxation to $(\epsilon,\delta)$-privacy, as results such as those given in this paper cannot be extended to arbitrary low-sensitivity queries while satisfying $\epsilon$-differential privacy as shown recently by De \cite{De11}.

Roth and Roughgarden \cite{RR10} showed that bounds similar to those achieved in this paper can also be achieved in the \emph{interactive} setting, in which queries are allowed to arrive online and must be answered before the next query is known. In many applications, this gives a large improvement in the accuracy of answers, because it allows the analyst to pay for those queries which were actually asked in the course of a computation (which may be only polynomially many), as opposed to all queries which might potentially be asked, as is necessary for a non-interactive mechanism. Hardt and Rothblum \cite{HR10} gave an improved mechanism for the interactive setting based on the multiplicative weights framework which achieves bounds comparable to the improved bounds of \cite{DRV10}, also in the interactive setting. An offline version of this mechanism (constructed by pairing the online mechanism with an agnostic learner for the class of queries of interest) was given by Gupta et al. \cite{GHRU11}, and an experimental evaluation on real data was done by Hardt, Ligett, and McSherry \cite{HLM11}. Gupta, Roth, and Ullman unified the online mechanisms of \cite{RR10,HR10} into a generic framework (and improved their error bounds) by giving a generic reduction from online learning algorithms in the mistake bound model to private query release algorithms in the interactive setting \cite{GRU11}. \cite{GRU11} also gives a new mechanism based on this reduction that achieves improved error guarantees for the setting in which the database size is comparable to the size of the data universe.

Following this work, there has been significant attention paid to the problem of releasing the class of conjunctions (a special case of counting queries) with low error using algorithms with more efficient run-time than the one given in this paper. Gupta et al \cite{GHRU11} give an algorithm which runs in time polynomial in the size of the database, and releases the class of conjunctions to $O(1)$ \emph{average} error while preserving differential privacy. Hardt, Rothblum, and Servedio \cite{HRS11} give an algorithm which runs in time proportional $d^k$ (for databases over a data universe $X = \{0,1\}^d)$ and releases conjunctions of most $k$ variables with worst-case error guarantees. Their algorithm improves over the Laplace mechanism (which also requires run-time $d^k$) because it only requires that the database size be proportional to $d^{\sqrt{k}}$ (The Laplace mechanism would require a database of size $d^k$). As a building block for this result, they also give a mechanism with run-time proportional to $d^{\sqrt{k}}$ which gives average-case error guarantees. Kasiviswanathan et al. \cite{KRSU10} extend the lower bounds \cite{DN03} from arbitrary subset-sum queries to hold also for an algorithm that only releases conjunctions.

Xiao et al. \cite{XWG10} gave an algorithm for releasing range queries, which extends the class of constant-dimensional interval queries which we consider in this paper.

There has also been progress in proving lower bounds. Dwork et al. \cite{DNRRV09} show that in general, the problem of releasing synthetic data giving non-trivial error for arbitrary classes of counting queries requires run-time that is linear in the size of the data universe and the size of the query class (modulo cryptographic assumptions). This in particular precludes improving the run-time of the general mechanism presented in this paper to be only polynomial in the size of the database. Ullman and Vadhan \cite{UV11} extend this result to show that releasing synthetic data is hard even for the simple class of conjunctions of at most 2 variables. This striking result emphasizes that output representation is extremely important, because it is possible to release the answers to all of the (at most $d^2$) conjunctions of size 2 privately and efficiently using output representations other than synthetic data. Hardt and Talwar showed how to prove lower bounds for differentially query release using packing arguments, and gave an optimal lower bound for a certain range of parameters \cite{HT10}. De recently refined this style of argument and extended it to wider settings \cite{De11}. Gupta et al. \cite{GHRU11} showed that the class of queries that can be released by mechanisms that access the database using only \emph{statistical queries} (which includes almost all mechanisms known to date, with the exception of the parity learning algorithm of \cite{KLNRS08}) is equal to the class of queries that can be agnostically learned using statistical queries. This rules out a mechanism even for releasing conjunctions to subconstant error which accesses the data using only a polynomial number of statistical queries.

{
\subsection{Motivation from Learning Theory}
From a machine learning perspective, one of the main {\em reasons} one would
want to perform statistical analysis of a database in the first place
is to gain information about the population from which that database
was drawn.  In particular, a fundamental result in learning theory is
that if one views a database as a
collection of random draws from some distribution $\D$, and one is
interested in a particular class $C$ of boolean predicates over
examples, then a database $D$ of size
$\tilde{O}(\textrm{VCDIM}(C)/\alpha^2)$ is
sufficient so that with high probability, for {\em every} query $q \in
C$, the proportion of examples in $D$ satisfying $q$ is within $\pm
\alpha$ of the true probability mass under $\D$
\cite{AB99,Vapnik:book98}.\footnote{{Usually,
this kind of uniform convergence is stated as empirical error
approximating true error.  In our setting, we have no notion of an
``intrinsic label'' of database elements.  Rather, we imagine that
different users may be interested in learning different things.  For
example, one user might want to learn a rule to predict feature $x_d$
from features $x_1, \ldots, x_{d-1}$; another might want to use the
first half of the features to predict a certain boolean function over
the second half.}}  Our main result can be viewed as asking how much
larger does a database $D$ have to be in order to do this in a
privacy-preserving manner: that is, to allow one to
(probabilistically) construct an output $\hat{D}$ that accurately
approximates $\D$ with respect to all queries in $C$, and yet that
reveals no extra information about database $D$.\footnote{{Formally, we
only care about $\hat{D}$ approximating $D$ with respect to $C$, and
want this to be true no matter how $D$ was constructed.  However, if
$D$ {\em was} a random sample from a distribution $\D$, then $D$ will
approximate $\D$ and therefore $\hat{D}$ will as well.}}  In fact, our
notion of distributional privacy (Section \ref{sec:distributional}) is
motivated by this view.  Note that since {\em interactive} privacy
mechanisms can handle arbitrary queries of this form so long as only
$o(n)$ are requested, our objective is interesting only for classes
$C$ that contain $\Omega(n)$, or even exponentially in $n$ many
queries.  We will indeed achieve this (Theorem \ref{polyVC}), since
$|C| \geq 2^{\textrm{VCDIM(C)}}$.}

\vspace{.75cm}
\subsection{Organization}
We present essential definitions in Section \ref{sec:definitions}.  In
Section \ref{sec:VC}, we show that, ignoring computational constraints,
one can release sanitized databases over discretized domains that are
useful for {\em{any}} concept class with polynomial VC-dimension.  We
then, in Section \ref{sec:interval}, give an efficient algorithm for
privately releasing a database useful for the class of interval queries.
We next turn to the study of halfspace queries over $\bR^d$ and show in
Section \ref{sec:halfspaceLB} that, without relaxing the definition of
usefulness, one cannot release a database that is privacy-preserving and
useful for halfspace queries over a continuous domain.  Relaxing our
definition of usefulness, in Section \ref{sec:algorithm}, we give an
algorithm that in polynomial time, creates a sanitized database that
usefully and privately answers all halfspace queries.  We present an
alternative definition of privacy and discuss its relationship to
differential privacy in Section \ref{sec:distributional}.

\section{Definitions}\label{sec:definitions}
We consider databases which are $n$-tuples from some abstract domain $X$: i.e. $D \in X^n$. We will write $n = |D|$ for the size of the database. We think of $X$ as the set of all possible data-records. For example, if data elements are represented as bit-strings of length $d$, then $X = \{0,1\}^d$ would be the boolean hypercube in $d$ dimensions. These tuples are not endowed with an ordering: they are simply multi-sets (they can contain multiple copies of the same element $x \in X$).

A database access mechanism is a randomized mapping $A:X^*\rightarrow R$, where $R$ is some arbitrary range. We say that $A$ outputs synthetic data if its output is itself a database: i.e. if $R = X^*$.
%
%For a database $D$, let $A$ be a database access
%mechanism. For an interactive mechanism, we will say that $A(D,Q)$
%induces a distribution over outputs for each query $Q$. For a
%non-interactive mechanism, we will say that $A(D)$ induces a
%distribution over outputs.

Our privacy solution concept will be the by now standard notion of differential privacy. Crucial to this definition will be the notion of \emph{neighboring databases}. We say that two databases $D, D' \in X^n$ are \emph{neighboring} if they differ in only a single data element: i.e. they are neighbors if their symmetric difference $|D \Delta D'| \leq 1$.
\begin{definition}[Differential Privacy \cite{DMNS06}]
A database access mechanism $A:X^n\rightarrow R$ is $\epsilon$-differentially private if for all neighboring pairs of databases $D, D' \in X^n$ and for all outcome events $S \subseteq R$, the following holds:
$$\Pr[A(D) \in S] \leq \exp(\epsilon)\Pr[A(D') \in S]$$
\end{definition}

%We say that an interactive database access mechanism $A$ satisfies
%{\em {$\epsilon$-differential privacy}} if for all neighboring databases $D_1$
%and $D_2$ (differing in only a single element), for all queries $Q$,
%and for all outputs $x$,
%$$\Pr[A(D_1,Q) = x] \leq e^{\epsilon}\Pr[A(D_2,Q)=x].$$ We say that a
%non-interactive database sanitization mechanism $A$ satisfies
%{\em{$\epsilon$-differential privacy}} if for all neighboring databases $D_1$
%and $D_2$, and for all sanitized outputs $\widehat{D}$,
%$$\Pr[A(D_1)=\widehat{D}] \leq e^\epsilon\Pr[A(D_2) = \widehat{D}].$$

In Section \ref{sec:distributional}, we propose an alternate
definition of privacy, distributional privacy, and show that it is
strictly stronger than differential privacy.  For simplicity, however,
in the main body of the paper, we use the standard definition,
differential privacy.  All of these proofs can be adapted to the
distributional privacy notion.

\begin{definition}
The {\em{global sensitivity}} of a query $f$ is its maximum difference when
evaluated on two neighboring databases:
$$GS_f = \max_{D,D' \in X^n: |D\Delta D'| = 1} |f(D)-f(D')|.$$
\end{definition}

In this paper, we consider the private release of information useful for classes of \emph{counting queries}.

\begin{definition}
A {\em{counting query}} $Q_\varphi$, defined in terms of a predicate $\varphi:X\rightarrow \{0,1\}$ is defined to be
$$Q_\varphi(D) = \frac{\sum_{x \in D}\varphi(x)}{|D|}.$$ It evaluates to the fraction of elements in the database that satisfy the predicate $\varphi$.
\end{definition}
\begin{observation}
\label{predicateSensitivity}
For any predicate $\varphi:X\rightarrow \{0,1\}$, the corresponding counting query $Q_\varphi:X^n\rightarrow [0,1]$ has global sensitivity $GS_{Q_\varphi} \leq 1/n$
\end{observation}
\begin{proof}
Let $D,D' \in X^n$ be neighboring databases. Then:
$$Q_\varphi(D) =   \frac{|\{x\in D : \varphi(x) = 1\}|}{|D|} \leq \frac{|\{x\in D' : \varphi(x) = 1\}| + 1}{|D|} =  Q_\varphi(D') + 1/n$$
Where the inequality follows by the definition of neighboring databases. Similarly, $Q_\varphi(D) \geq Q_\varphi(D') - 1/n$. The observation then follows.
\end{proof}

We remark that everything in this paper easily extends to the case of more general \emph{linear} queries, which can are defined analogously to counting queries, but involve real valued predicates $\varphi:X\rightarrow [0,1]$. For simplicity we restrict ourselves to counting queries in this paper, but see \cite{Rot10} for the natural extension to linear queries.

A key measure of complexity that we will use for counting queries is VC-dimension. VC-dimension is strictly speaking a measure of complexity of classes of predicates, but we will associate the VC-dimension of classes of predicates with their corresponding class of counting queries.
\begin{definition}[Shattering]
A class of predicates $P$ \emph{shatters} a collection of points $S \subseteq X$ if for every $T \subseteq S$, there exists an $\varphi \in P$ such that $\{x \in S : \varphi(x) = 1\} = T$. That is, $P$ shatters $S$ if for every one of the $2^{|S|}$ subsets $T$ of $S$, there is some predicate in $P$ that labels exactly those elements as positive, and does not label any of the elements in $S \setminus T$ as positive.
\end{definition}
We can now define our complexity measure for counting queries.
\begin{definition}[VC-Dimension]
A collection of predicates $P$ has VC-dimension $d$ if there exists some set $S \subseteq X$ of cardinality $|S| = d$ such that $P$ shatters $S$, and $P$ does not shatter any set of cardinality $d+1$. We can denote this quantity by $\textrm{VC-DIM}(P)$. We abuse notation and also write $\textrm{VC-DIM}(C)$ where $C$ is a class of counting queries, to denote the VC-dimension of the corresponding collection of predicates.
\end{definition}

Dwork et al. \cite{DMNS06} give a mechanism which can answer any single low-sensitivity query while preserving differential
privacy:
\begin{definition}[Laplace mechanism]
The Laplace mechanism
responds to a query $Q$ by returning $Q(D) + Z$ where $Z$ is a random variable drawn from the Laplace distribution: $Z \sim \textrm{Lap}(GS_Q/\epsilon)$.

The Laplace distribution with scale $b$, which we denote by $\textrm{Lap}(b)$, has probability density function
 $$f(x|b) = \frac{1}{2b}\exp\left(-\frac{|x|}{b}\right)$$
\end{definition}

\begin{theorem}[Dwork et al.\ \cite{DMNS06}]
The Laplace mechanism preserves $\epsilon$-differential privacy.
\end{theorem}

%However, lower bounds of Dinur and Nissim \cite{DN03} and Dwork et
%al.\ \cite{DMNS06} imply that such mechanisms can only answer a
%sublinear number of queries on any database. Note that these
%mechanisms can only answer a sublinear number of queries \textit{in
%  total}, not per user.
This mechanism answers queries interactively, but for a fixed privacy level, its accuracy guarantees degrade linearly in the number of queries that it answers. The following composition theorem is useful: it tells us that a mechanism which runs $k$ $\epsilon$-differentially private subroutines is $k\epsilon$-differentially private.

 \begin{theorem}[Dwork et al.\ \cite{DMNS06}]
 If mechanisms $M_1,\ldots,M_k$ are each $\epsilon$-differentially private, then the mechanism $M$ defined by the (string) composition of the $k$ mechanisms: $M(D) = (M_1(D),\ldots,M_k(D))$ is $k\epsilon$-differentially private.
 \end{theorem}

  We propose to construct database access mechanisms which produce one-shot (non-interactive) outputs that can be
released to the public, and so can necessarily be used to answer an
arbitrarily large number of queries. We seek to do this while
simultaneously preserving privacy.  However, as implied by the lower
bounds of Dinur and Nissim \cite{DN03}, we cannot hope to be able to usefully answer arbitrary
queries. We instead seek to release synthetic databases which are ``useful'' (defined below) for restricted classes of queries $C$.

\begin{definition}[Usefulness]
A database access mechanism $A$ is {\em{$(\alpha,\delta)$-useful}} with respect to queries in
class $C$ if for every database $D \in X^n$, with probability at least $1-\delta$, the output of the mechanism $\widehat{D} = A(D)$ satisfies:
$$\max_{Q \in C}|Q(\widehat{D}) - Q(D)| \leq \alpha$$
\end{definition}

In this paper, we will derive $(\alpha,\delta)$-useful mechanisms from small $\alpha$-nets:
\begin{definition}[$\alpha$-net]
An $\alpha$-net of databases with respect to a class of queries $C$ is a set $N \subset X^*$ such that for all $D \in X^*$, there exists an element of the $\alpha$-net $D' \in N$ such that:
$$\max_{Q \in C}|Q(D)-Q(D')| \leq \alpha$$
We write $N_\alpha(C)$ to denote an $\alpha$-net of minimum cardinality among the set of all $\alpha$-nets for $C$.
\end{definition}

\section{General release mechanism}\label{sec:VC}
In this section we present our general release mechanism. It is an instantiation of the \emph{exponential mechanism} of McSherry and Talwar \cite{MT07}.

 Given some arbitrary range $\mathcal{R}$, the exponential mechanism is defined with respect so some quality function $q:X^*\times\mathcal{R}\rightarrow \mathbb{R}$, which maps database/output pairs to quality scores. We should interpret this intuitively as a measure stating that fixing a database $D$, the user would prefer the mechanism to output some element of $\mathcal{R}$ with as high a quality score as possible.

\begin{definition}[The Exponential Mechanism \cite{MT07}]
The exponential mechanism $M_E(D, q, \mathcal{R})$ selects and outputs an element $r \in \mathcal{R}$ with probability proportional to $\exp(\frac{\eps q(D,r)}{2GS_q})$.
\end{definition}
McSherry and Talwar showed that the exponential mechanism preserves differential privacy. It is important to note that the exponential mechanism can define a complex distribution over a large arbitrary domain, and so it may not be possible to implement the exponential mechanism efficiently when the range of $q$ is super-polynomially large in the natural parameters of the problem. This will be the case with our instantiation of it.
\begin{theorem}[\cite{MT07}]
\label{thm:exponential-mechanism-privacy}
The exponential mechanism preserves $\eps$-differential privacy.
\end{theorem}

\begin{algorithm}
\label{alg:alphaNet}
\caption{The Net Mechanism}
\textbf{NetMechanism}($D,C,\epsilon,\alpha$)
\begin{algorithmic}
\STATE \textbf{Let} $\mathcal{R} \leftarrow N_\alpha(C)$
\STATE \textbf{Let} $q:X^*\times\mathcal{R} \rightarrow \mathbb{R}$ be defined to be:
$$q(D,D') = - \max_{Q \in C}\left|Q(D)-Q(D')\right|$$
\STATE \textbf{Sample And Output} $D' \in \mathcal{R}$ with the exponential mechanism $M_E(D,q,\mathcal{R})$
\end{algorithmic}
\end{algorithm}
We first observe that the Net mechanism preserves $\eps$-differential privacy.
\begin{proposition}
The Net mechanism is $\epsilon$-differentially private.
\end{proposition}
\begin{proof}
The Net mechanism is simply an instantiation of the exponential mechanism. Therefore, privacy follows from Theorem \ref{thm:exponential-mechanism-privacy}.
\end{proof}
We may now analyze the usefulness of the net mechanism.
\begin{proposition}
\label{prop:BLR-utility}
For any class of queries $C$ (not necessarily counting queries) the Net Mechanism is $(2\alpha,\delta)$-useful for any $\alpha$ such that:
$$\alpha \geq \frac{2\Delta}{\epsilon}\log\frac{N_\alpha(C)}{\delta}$$
Where $\Delta = \max_{Q \in C}GS_Q$.
\end{proposition}
\begin{proof}
First observe that the sensitivity of the quality score $GS_q \leq \max_{Q \in C} GS_Q = \Delta$.

By the definition of an $\alpha$-net, we know that there exists some $D^* \in \mathcal{R}$ such that $q(D,D^*) \geq -\alpha$. By the definition of the exponential mechanism, this $D^*$ is output with probability proportional to at least $\exp(\frac{-\eps \alpha}{2GS_q})$. Similarly, there are at most $|N_\alpha(C)|$ databases $D' \in \mathcal{R}$ such that $q(D,D') \leq -2\alpha$ (simply because $\mathcal{R} = N_\alpha(C)$). Hence, by a union bound, the probability that the exponential mechanism outputs some $D'$ with $q(D,D') \leq -2\alpha$ is at most $|N_\alpha(C)|\exp(\frac{-2\eps \alpha}{2GS_q})$. Therefore, if we denote by $A$ the event that the net mechanism outputs some $D^*$ with $q(D,D^*) \geq -\alpha$, and denote by $B$ the event that the net mechanism outputs some $D'$ with $q(D,D') \leq -2\alpha$, we have:
\begin{eqnarray*}
\frac{\Pr[A]}{\Pr[B]} &\geq& \frac{\exp(\frac{-\eps \alpha}{2\Delta})}{|N_\alpha(C)|\exp(\frac{-2\eps \alpha}{2\Delta})} \\
&=& \frac{\exp(\frac{\eps\alpha}{2\Delta})}{|N_\alpha(C)|}
\end{eqnarray*}
Note that if this ratio is at least $1/\delta$, then we will have proven that the net mechanism is $(2\alpha,\delta)$ useful with respect to $C$. Solving for $\alpha$, we find that this is condition is satisfied so long as $$\alpha \geq \frac{2\Delta}{\epsilon}\log\frac{N_\alpha(C)}{\delta}$$
\end{proof}

We have therefore reduced the problem of giving upper bounds on the usefulness of differentially private database access mechanisms to the problem of upper bounding the sensitivity of the queries in question, and the size of the smallest $\alpha$-net for the set of queries in question. Recall that for \emph{counting} queries $Q$ on databases of size $n$, we always have $GS_Q \leq 1/n$.  Therefore we have the immediate corollary:
\begin{corollary}
\label{cor:utility}
For any class of counting queries $C$ the Net Mechanism is $(2\alpha,\delta)$-useful for any $\alpha$ such that:
$$\alpha \geq \frac{2}{\epsilon n}\log\frac{N_\alpha(C)}{\delta}$$
\end{corollary}

To complete the proof of utility for the net mechanism for counting queries, it remains to prove upper bounds on the size of minimal $\alpha$-nets for counting queries. We begin with a bound for finite classes of queries.

\begin{theorem}
\label{thm:alpha-net-bound}
For any finite class of counting queries $C$:
$$\left|N_\alpha(C)\right| \leq \left|X\right|^{\frac{\log|C|}{\alpha^2}}$$
\end{theorem}
In order to prove this theorem, we will show that for any collection of counting queries $C$ and for any database $D$, there is a ``small'' database $D'$ of size $|D'| = \frac{\log|C|}{\alpha^2}$ that approximately encodes the answers to every query in $C$, up to error $\alpha$. Crucially, this bound will be independent of $|D|$.
\begin{lemma}
\label{lem:alpha-net-bound}
For any $D \in X^*$ and for any finite collection of counting queries $C$, there exists a database $D'$ of size
$$|D'| = \frac{\log|C|}{\alpha^2}$$
such that:
$$\max_{Q\in C}\left|Q(D)-Q(D')\right| \leq \alpha$$
\end{lemma}
\begin{proof}
Let $m = \frac{\log|C|}{\alpha^2}$ We will construct a database $D'$ by taking $m$ uniformly random samples from the elements of $D$. Specifically, for $i \in \{1,\ldots, m\}$ let $X_i$ be a random variable taking value $x_j$ with probability $|\{x \in D : x = x_j\}|/|D|$, and let $D'$ be the database containing elements $X_1,\ldots,X_m$. Now fix any $Q_\varphi \in C$ and consider the quantity $Q_\varphi(D')$. We have:
$Q_\varphi(D') = \frac{1}{m}\sum_{i=1}^m\varphi(X_i)$.
We note that each term of the sum $\varphi(X_i)$ is a bounded random variable taking values $0 \leq \varphi(X_i) \leq 1$, and that the expectation of $Q_\varphi(D')$ is:
$$\E[Q(D')] = \frac{1}{m} \sum_{i=1}^m \E[\varphi(X_i)] = Q_\varphi(D)$$
Therefore, we can apply a standard Chernoff bound which gives:
$$\Pr\left[\left|Q_\varphi(D') - Q_\varphi(D)\right| > \alpha\right] \leq 2e^{-2m\alpha^2}$$
Taking a union bound over all of the counting queries $Q_\varphi \in C$, we get:
$$\Pr\left[\max_{Q_\varphi \in C}\left|Q_\varphi(D') - Q_\varphi(D)\right| > \alpha\right] \leq 2|C|e^{-2m\alpha^2}$$
Plugging in $m$ makes the right hand side smaller than $1$ (so long as $|C| > 2$), proving that there exists a database of size $m$ satisfying the stated bound, which completes the proof of the lemma.
\end{proof}
Now we can complete the proof of Theorem \ref{thm:alpha-net-bound}.
\begin{proof}[Proof of Theorem \ref{thm:alpha-net-bound}]
By Lemma \ref{lem:alpha-net-bound}, we have that for any $D \in X^*$ there exists a database $D' \in X^*$ with $|D'| = \frac{\log|C|}{\alpha^2}$ such that $\max_{Q_\varphi\in C}\left|Q_\varphi(D)-Q_\varphi(D')\right| \leq \alpha$. Therefore, if we take $N = \{D' \in X^* : |D'| = \frac{\log|C|}{\alpha^2}\}$ to be the set of \emph{every} database of size $\frac{\log|C|}{\alpha^2}$, we have an $\alpha$-net for $C$. Since
$$\left|N\right| =  \left|X\right|^{\frac{\log|C|}{\alpha^2}}$$
and by definition $\left|N_\alpha(C)\right| \leq \left|N\right|$, we have proven the theorem.
\end{proof}

For infinite concept classes $C$, we can replace lemma \ref{lem:alpha-net-bound} with the following lemma:
\begin{lemma}[\cite{AB99,Vapnik:book98}]
\label{epsilonCover}
For any $D \in X^*$ and for any collection of counting queries $C$, there exists a database $D'$ of size
$$|D'| = O(\textrm{VCDIM}(C){\log(1/\alpha)}/\alpha^2)$$
such that:
$$\max_{Q\in C}\left|Q(D)-Q(D')\right| \leq \alpha$$
\end{lemma}

This lemma straightforwardly gives an analogue of Theorem \ref{thm:alpha-net-bound}:
\begin{theorem}
\label{thm:vc-alpha-net-bound}
For any class of counting queries $C$:
$$\left|N_\alpha(C)\right| \leq \left|X\right|^{O(\textrm{VCDIM}(C){\log(1/\alpha)}/\alpha^2)}$$
\end{theorem}

Note that we always have $\textrm{VCDIM}(C) \leq \log |C|$ for finite classes of counting queries, and so Theorem \ref{thm:vc-alpha-net-bound} is strictly stronger than Theorem \ref{thm:alpha-net-bound}.

Finally, we can instantiate Corollary \ref{cor:utility} to give our main utility theorem for the Net Mechanism.
\begin{theorem}
\label{polyVC}
For any class of counting queries $C$ the Net Mechanism is $(\alpha,\delta)$-useful for any $\alpha$ such that:
$$\alpha \geq O\left(\frac{1}{\epsilon \alpha^2 n}\left(\textrm{VCDIM}(C)\log |X|\log(1/\alpha)+ \log 1/\delta\right)\right)$$

Solving for $\alpha$, the Net Mechanism is $(\alpha,\delta)$-useful for:
$$\alpha = \tilde{O}\left(\left(\frac{\textrm{VCDIM}(C)\log X + \log 1/\delta}{\epsilon n}\right)^{1/3}\right)$$
\end{theorem}

Theorem \ref{polyVC} shows that a database of size
$\widetilde{O}(\frac{\log X\textrm{VCDIM}(C)}{\alpha^3\epsilon})$ is
sufficient in order to output a set of points that is
$\alpha$-useful for a concept class $C$, while simultaneously
preserving $\epsilon$-differential privacy. If we were to view our
database as having been drawn from some distribution $\mathcal{D}$, this
is only an extra $\widetilde{O}(\frac{\log X}{\alpha\epsilon})$ factor larger
than what would be required to achieve $\alpha$-usefulness with
respect to $\mathcal{D}$, even without any privacy guarantee!

The results in this section only
apply for discretized database domains, and
may not be computationally efficient.  We explore these two issues
further in the remaining sections of the paper.

\subsection{The Necessity of a Dependence on VC-Dimension}
We just gave an $\epsilon$-differentially private mechanism that is $(\alpha,\delta)$-useful with respect to any set of counting queries $C$, when given a database of size $n \geq \widetilde{O}(\frac{\log X\textrm{VCDIM}(C)}{\alpha^3\epsilon})$. In this section, we show that the dependence on the VC-dimension of the class $C$ is tight. Namely:

\begin{theorem}
For any class of counting queries $C$, for any $0 < \delta < 1$ bounded away from $1$ by a constant, for any $\epsilon \leq 1$, if $M$ is an $\epsilon$-differentially private mechanism that is $(\alpha,\delta)$ useful for $C$ given databases of size $n \leq \frac{\textrm{VCDIM}(C)}{2}$, then $\alpha \geq \Omega\left(\frac{1}{4+16\epsilon}\right)$.
\end{theorem}
\begin{proof}
Fix a class of counting queries $C$ corresponding to a class of predicates $P$ of VC-dimension $d$. Let $S \subset X$ denote a set of universe elements of size $|S| = d$ that are \emph{shattered} by $P$, as guaranteed by the definition of VC-dimension. We will consider all subsets $T \subset S$ of size $|T| = d/2$. Denote this set by $\mathcal{D}_S = \{T \subset S : |T| = d/2\}$ For each such $T \in \mathcal{D}_S$, let $\varphi_T$ be the predicate such that:
$$\varphi_T(x) = \left\{
                                                                                                                                                                          \begin{array}{ll}
                                                                                                                                                                            1, & \hbox{$x\in T$;} \\
                                                                                                                                                                            0, & \hbox{$x\not\in T$.}
                                                                                                                                                                          \end{array}
                                                                                                                                                                        \right.$$
as guaranteed by the definition of shattering, and let $Q_T = Q_{\varphi_T}$ be the corresponding counting query.

First we prove a simple lemma:
\begin{lemma}
\label{lem:symdiff}
For every pair $T, T' \in \mathcal{D}_S$:
$$Q_T(T) - Q_T(T') = \frac{|T\Delta T'|}{d}$$
\end{lemma}
\begin{proof}
\begin{eqnarray*}
Q_T(T) - Q_T(T') &=& \frac{1}{|T|}\left(\sum_{x \in T} \varphi_T(x) - \sum_{x \in T'} \varphi_T(x) \right)\\
&=& \frac{2}{d}\left(\sum_{x \in T \cap T'}(\varphi_T(x) - \varphi_T(x)) + \sum_{x \in T \setminus T'}\varphi_T(x) -  \sum_{x \in T' \setminus T}\varphi_T(x)\right) \\
&=& \frac{2}{d}|T\setminus T'| \\
&=&  \frac{|T\Delta T'|}{d}
\end{eqnarray*}
where the last equality follows from the fact that $|T| = |T'|$.
\end{proof}

This lemma will allow a sufficiently useful mechanism for the class of queries $C$ to be used to reconstruct a database selected from $\mathcal{D}_S$ with high accuracy.
\begin{lemma}
\label{lem:reconstruct}
For any $0 < \delta < 1$ bounded away from $1$ by a constant, let $M$ be an $(\alpha,\delta)$-useful mechanism for $C$. Given as input $M(T)$ where $T$ is any database $T \in \mathcal{D}_S$, there is a procedure which with constant probability $1-\delta$ reconstructs a database $T'$ with $|T' \Delta T| \leq 2d\alpha$.
\end{lemma}
\begin{proof}
Write $D' = M(T)$. With probability at least $1-\delta$, we have $\max_{Q \in C}|Q(T)-Q(D')| \leq \alpha$. For the rest of the argument, assume this event occurs. For each $T' \in \mathcal{D}_S$, define:
$$v_{T'} = Q_{T'}(T') - Q_{T'}(D')$$
Let $T' = \argmin_{T' \in \mathcal{D}_S}v_{T'}$. We have:
$$v_{T'} \leq v_T = Q_{T}(T) - Q_{T}(D') \leq \alpha$$
by the accuracy of the mechanism. By combining the accuracy of the mechanism with Lemma \ref{lem:symdiff}, We also have:
$$v_{T'} \geq \frac{|T\Delta T'|}{d} - \alpha$$
Combining these two inequalities yields a database $T'$ such that: $|T\Delta T'| \leq 2d\alpha$ as claimed.
\end{proof}

We can now complete the proof. Let $T \in \mathcal{D}_S$ be a set selected uniformly at random, let $x \in T$  be an element of $T$ selected uniformly at random, and let $y \in S \setminus T$ be an element of $S\setminus T$ selected uniformly at random. Note that the marginal distributions on $x$ and $y$ are identical: both are uniformly random elements from $S$. Let $\hat{T} = (T\setminus \{x\})\cup \{y\}$ be the set obtained by swapping elements $x$ and $y$. Note that $\hat{T}$ is also distributed uniformly at random from $\mathcal{D}_S$. Let $T'$ be the set reconstructed from $D' = M(T)$ as in lemma \ref{lem:reconstruct}, and let $\hat{T}'$ be the set reconstructed from $D' = M(\hat{T})$. Assume that $|T \Delta T'| \leq 2d\alpha$ and that $|\hat{T} \Delta \hat{T}'| \leq 2d\alpha$, which occurs except with probability at most $2\delta$. We now have:
$$\Pr[x \in T'] = \frac{|T| - (1/2)|T \Delta T'|}{|T|} \geq \frac{\frac{d}{2} - d\alpha}{\frac{d}{2}} = 1 - 2\alpha$$
By symmetry:
$$\Pr[x \in \hat{T}'] = \frac{(1/2)|\hat{T} \Delta \hat{T}'|}{|\hat{T}|} \leq \frac{d\alpha}{\frac{d}{2}} = 2\alpha$$
Now recall that $|T \Delta \hat{T}| \leq 2$ and so by the fact that $M$ is $\epsilon$-differentially private, we also know:
$$\exp(2\epsilon) \geq\frac{\Pr[x \in T']}{\Pr[x \in \hat{T}']} \geq \frac{1-2\alpha}{2\alpha} = \frac{1}{2\alpha}-1$$
Because we also have $\exp(2\epsilon) \leq 1+8\epsilon$, we can solve to find $\alpha \geq \frac{1}{4+16\epsilon}$.
\end{proof}

\section{Interval queries}\label{sec:interval}
In this section we give an {\em {efficient}} algorithm for privately
releasing a database useful for the class of interval queries over a
discretized domain, given a database of size only polynomial in our
privacy and usefulness parameters. We note that our algorithm is easily
extended to the class of axis-aligned rectangles in $d$ dimensional
space for $d$ a constant; we present the case of $d=1$ for clarity.

Consider a database $D$ of $n$ points in a discrete interval $\{1,\ldots,2^d\}$ (in Corollary \ref{impossibility} we show some discretization is
necessary). Given $a_1 \leq a_2$,  both in $\{1,2,\ldots,2^d\}$, let $I_{a_1,a_2}$ be the indicator
function corresponding to the interval $[a_1,a_2]$. That is:
$$I_{a_1,a_2}(x) = \left\{
                 \begin{array}{ll}
                   1, & \hbox{$a_1 \leq x \leq a_2$;} \\
                   0, & \hbox{otherwise.}
                 \end{array}
               \right.$$

\begin{definition}
An interval query $Q_{[a_1,a_2]}$ is defined to be
$$Q_{[a_1,a_2]}(D) = \sum_{x \in D}\frac{I_{a_1,a_2}(x)}{|D|}.$$
\end{definition}

Note that $GS_{Q_{[a_1,a_2]}} = 1/n$, and we may answer interval queries
while preserving $\epsilon$-differential privacy by adding noise
proportional to $\textrm{Lap}(1/(\epsilon n))$.

We now give the algorithm. We will assume for simplicity that all points $x, x' \in D$ are distinct, but this condition can be easily discarded.
\begin{algorithm}
\label{alg:intervals}
\caption{An algorithm for releasing synthetic data for interval queries.}
\textbf{ReleaseIntervals}$(D, \alpha, \eps)$
\begin{algorithmic}
\STATE \textbf{Let} $\alpha' \leftarrow \alpha/6$, $\textrm{MaxIntervals} \leftarrow \lceil 4/3\alpha'\rceil$, $\epsilon' \leftarrow  \eps/(d\cdot\textrm{MaxIntervals})$.
\STATE \textbf{Let} Bounds be an array of length $\textrm{MaxIntervals}$
\STATE \textbf{Let} $i \leftarrow 1$, Bounds$[0] \leftarrow 1$
\WHILE{Bounds$[i-1] < 2^d$}
    \STATE $a \leftarrow \textrm{Bounds}[i-1]$, $b \leftarrow (2^{d}-a+1)/2$, $\textrm{increment} \leftarrow (2^{d}-a+1)/4$
    \WHILE{$\textrm{increment} \geq 1$}
        \STATE \textbf{Let} $\hat{v} \leftarrow Q_{[a,b]}(D) + \textrm{Lap}(1/(\eps'n))$
        \IF{$\hat{v} > \alpha'$}
            \STATE \textbf{Let} $b \leftarrow b - \textrm{increment}$
        \ELSE
            \STATE \textbf{Let} $b \leftarrow b + \textrm{increment}$
        \ENDIF
        \STATE \textbf{Let} $\textrm{increment} \leftarrow \textrm{increment}/2$
    \ENDWHILE
    \STATE \textbf{Let} Bounds$[i] \leftarrow b$, $i \leftarrow i+1$
\ENDWHILE
\STATE \textbf{Output} $D'$, a database that has $\alpha' m$ points in each interval $[$Bounds$[j-1]$, Bounds$[j]]$ for each $j \in [i]$, for any $m > \frac{1}{\alpha'}$.
\end{algorithmic}
\end{algorithm}
The algorithm \ref{alg:intervals} is very simple. It repeatedly performs a binary search to partition the unit interval into regions that have approximately an $\alpha'$ fraction of the point mass in them. It then releases a database that has exactly an $\alpha'$-fraction of the point mass in each of the intervals that it has discovered. There are at most $\approx 1/\alpha'$ such intervals, and each binary search terminates after at most $d$ rounds (because the interval consists of at most $2^d$ points). Therefore, the algorithm requires only $\approx d/\alpha'$ accesses to the database, and each one is performed in a privacy preserving manner using noise from the Laplace mechanism. The privacy of the mechanism then follows immediately:

\begin{theorem}
ReleaseIntervals is $\epsilon$-differentially private.
\end{theorem}
\begin{proof}
The algorithm runs a binary search at most $\lceil 4/3\alpha'\rceil$ times. Each time, the search halts after $d$ queries to the database using the Laplace mechanism. Each query is $1/\eps'$-differentially private (the sensitivity of an interval query is $1/n$ since it is a counting query). Privacy then follows from the definition of $\epsilon$ and the fact that the composition of $k$ differentially private mechanisms is $k\epsilon$ differentially private.
\end{proof}

\begin{theorem}
ReleaseIntervals is $(\alpha,\delta)$-useful for databases of size:
$$n \geq \frac{8 d}{\epsilon\alpha}\cdot \log\left(\frac{8d}{\delta \alpha}\right)$$
\end{theorem}

\begin{proof}
By a union bound and the definition of the Laplace distribution, if the database size $n$ satisfies the hypothesis of the theorem, then except with probability at most $\delta$, none of the $(4/3)d/\alpha'$ draws from the Laplace distribution have magnitude greater than $\alpha'^2$. That is, at every step, we have $|\hat{v} - Q_{[a,b]}(D)| \leq \alpha'^2$ except with probability $\beta$.  Conditioned on this event occurring, for each interval $[$Bounds$[j-1]$,Bounds$[j]]$ for $j \in [i]$, $f_{\textrm{Bounds}[j-1],\textrm{Bounds}[j]}(D) \in [\alpha' - \alpha'^2,\alpha' + \alpha'^2]$. In the synthetic database $D'$ released, each such interval contains exactly an $\alpha'$ fraction of the database elements. We can now analyze the error incurred on any query when evaluated on the synthetic database instead of on the real database. Any interval $[$Bounds$[j-1]$,Bounds$[j]] \subset [a,b]$ will contribute error at most $\alpha'$ to the total, and any interval $[$Bounds$[j-1]$,Bounds$[j]] \not\subset [a,b]$ that also intersects with $[a,b]$ contributes error at most $(\alpha' + \alpha'^2)$ to the total. Note that there are at most 2 intervals of this second type. Therefore, on any query $Q_{[a,b]}$ we have:
\begin{eqnarray*}
|Q_{[a,b]}(D')-Q_{[a,b]}(D)| &\leq& \sum_{j : [\textrm{Bounds}[j-1],\textrm{Bounds}[j]] \cap [a,b] \neq \emptyset}|Q_{[\textrm{Bounds}[j-1],\textrm{Bounds}[j]]}(D) - Q_{[\textrm{Bounds}[j-1],\textrm{Bounds}[j]]}(D')| \\
%&\leq& \alpha'^2|\{j : \textrm{Bounds}[j-1],\textrm{Bounds}[j] \subset [a,b]\}| + (\alpha'+\alpha'^2) \cdot |\{j : \textrm{Bounds}[j-1],\textrm{Bounds}[j] \not\subset [a,b] \wedge [\textrm{Bounds}[j-1]\textrm{Bounds}[j]] \cap [a,b] \neq \emptyset\}| \\
&\leq& \frac{4}{3\alpha'}\alpha'^2 + 2(\alpha' + \alpha'^2) \\
&\leq& 6\alpha' \\
&=& \alpha
\end{eqnarray*}
\end{proof}

We note that although the class of intervals (and more generally, low
dimensional axis-aligned rectangles) is a simple class of functions,
it nevertheless contains exponentially many queries, and so
it is not feasible to simply ask all possible interval queries using
an interactive mechanism.

\section{Lower bounds}\label{sec:halfspaceLB}
Could we possibly modify the results of Sections \ref{sec:interval} and
\ref{sec:VC} to hold for non-discretized databases?  Suppose we could
usefully answer an arbitrary number of queries in some simple concept
class $C$ representing interval queries on the real line (for example,
``How many points are contained within the following interval?'')  while
still preserving privacy. Then, for any database containing
single-dimensional real valued points, we would be able to answer median
queries with values that fall between the $50-\delta,50+\delta$
percentile of database points by performing a binary search on $D$ using
$A$ (where $\delta = \delta(\alpha)$ is some small constant depending
on the usefulness parameter $\alpha$). However, answering such queries
is impossible while guaranteeing differential privacy. Unfortunately,
this would seem to rule out usefully answering queries in simple concept
classes such as halfspaces and axis-aligned rectangles, that are
generalizations of intervals.

We say that a mechanism answers a median query $M$ usefully if it outputs a real value $r$ such that $r$ falls within the $50-\delta,50+\delta$ percentile of points in database $D$ for some $\delta < 50$.

\begin{theorem}
\label{median}
No mechanism $A$ can answer median queries $M$ with outputs that fall
between the $50-\delta,50+\delta$ percentile with positive probability on any
real valued database $D$, while still preserving
$\epsilon$-differential privacy, for $\delta < 50$ and any
$\epsilon$.
\end{theorem}
\begin{proof}
Consider real valued databases containing elements in the interval
$[0,1]$. Let $D_0 = (0,\ldots,0)$ be the database containing $n$ points
with value 0. Suppose $A$ can answer median queries usefully. Then we must have $\Pr[A(D_0,M) = 0] > 0$ since every point in $D_0$ is $0$. Since $[0,1]$ is a continuous interval, there must be some value $v \in [0,1]$ such
that $\Pr[A(D_0,M) = v] = 0$. Let $D_n = (v,\ldots,v)$ be the database
containing $n$ points with value $v$. We must have $\Pr[A(D_n,M) = v] >
0$. For $1 < i < n$, let $D_i =
(\underbrace{0,\ldots,0}_{n-i},\underbrace{v,\ldots,v}_i)$. Then we must
have for some $i$, $\Pr[A(D_i,M) = v] = 0$ but $\Pr[A(D_{i+1},M) = v] >
0$. But since $D_i$ and $D_{i+1}$ differ only in a single element, this
violates differential privacy.
\end{proof}

\begin{corollary}
\label{impossibility}
No mechanism can be $(\alpha,\delta)$-useful for the
class of interval queries, nor for any class $C$ that generalizes
interval queries to higher dimensions (for example, halfspaces, axis-aligned
rectangles, or spheres), while preserving
$\epsilon$-differential privacy, for any $\alpha,\delta < 1/2$ and
any $\epsilon \geq 0$.
\end{corollary}
\begin{proof}
Consider any real valued database containing elements in the interval
$[0,1]$. If $A$ is $(\alpha,\delta)$-useful for interval queries and
preserves differential privacy, then we can construct a mechanism $A'$
that can answer median queries usefully while preserving
differential privacy. By Theorem \ref{median}, this is impossible.  $A'$
simply computes $\widehat{D} = A(D)$, and performs binary search on
$\widehat{D}$ to find some interval $[0,a]$ that contains $n/2 \pm
\alpha n$ points. Privacy is preserved since we only access $D$ through
$A$, which by assumption preserves $\epsilon$-differential privacy. With positive
probability, all interval queries on $\widehat{D}$ are correct to within
$\pm \alpha$, and so the binary search can proceed. Since $\alpha < 1/2$, the result follows.
\end{proof}

We may get around the impossibility result of Corollary
\ref{impossibility} by relaxing our definitions.  One approach is to
discretize the database domain, as we do in Sections \ref{sec:VC} and
\ref{sec:interval}.  Another approach, which we take in Section
\ref{sec:algorithm}, is to relax our definition of usefulness.

\section{Answering Halfspace Queries}\label{sec:algorithm}
In this section, we give a non-interactive mechanism for releasing the answers to ``large-margin halfspace'' queries, defined over databases consisting of $n$ unit-length points in $\mathbb{R}^d$. The mechanism we give here will be different from the other mechanisms we have given in two respects. First, although it is a non-interactive mechanism, it will not output synthetic data, but instead another data structure representing the answers to its queries. Second, it will not offer a utility guarantee for all halfspace queries, but only those that have ``large margin'' with respect to the private database. Large margin, which we define below, is a property that a halfspace has with respect to a particular database. Note that by our impossibility result in the previous section, we know that without a relaxation of our utility goal, no private useful mechanism is possible.

\begin{definition}[Halfspace Queries]
For a unit vector $y \in \mathbb{R}^d$, the \emph{halfspace} query $f_y:\mathbb{R}^d\rightarrow \{0,1\}$ is defined to be:
$$f_y(x) = \left\{
             \begin{array}{ll}
               1, & \hbox{If $\langle x, y \rangle > 0$;} \\
               0, & \hbox{Otherwise.}
             \end{array}
           \right.$$
Let $C_{H} = \{f_y : y \in \mathbb{R}^d, ||y||_2 = 1\}$ denote the set of all halfspace queries.
\end{definition}
With respect to a database, a halfspace can have a certain \emph{margin} $\gamma$:
\begin{definition}[Margin]
A halfspace query $f_y$ has margin $\gamma$ with respect to a database $D \in (\mathbb{R}^d)^n$ if for all $x \in D$: $|\langle x, y \rangle | \geq \gamma$.
\end{definition}

Before we present the algorithm, we will introduce a useful fact about random projections, called the Johnson-Lindenstrauss lemma. It states, roughly, that the norm of a vector is accurately preserved with high probability when the vector is projected into a lower dimensional space with a random linear projection.
\begin{theorem}[The Johnson-Lindenstrauss Lemma \cite{DG99,BBV06}]
\label{thm:JL}
For $d > 0$ an integer and any $0 < \varsigma, \tau < 1/2$, let $A$ be a $T\times d$ random matrix with $\pm 1/\sqrt{T}$ random entries, for $T \geq 20\varsigma^{-2}\log(1/\tau)$. Then for any $x \in \mathbb{R}^d$:
$$\Pr_A[|||Ax||_2^2 - ||x||_2^2| \geq \varsigma ||x||_2^2] \leq \tau$$
\end{theorem}
For our purposes, the relevant fact will be that norm preserving projections also preserve pairwise inner products with high probability. The following corollary is well known.
\begin{corollary}[The Johnson-Lindenstrauss Lemma for Inner Products]
\label{cor:JL}
For $d > 0$ an integer and any $0 < \varsigma, \tau < 1/2$, let $A$ be a $T\times d$ random matrix with $\pm 1/\sqrt{T}$ random entries, for $T \geq 20\varsigma^{-2}\log(1/\tau)$. Then for any $x \in \mathbb{R}^d$:
$$\Pr_A[|\langle(Ax),(Ay)\rangle - \langle x, y\rangle | \geq \frac{\varsigma}{2} (||x||_2^2+||y||_2^2)] \leq 2\tau$$
\end{corollary}
\begin{proof}
Consider the two vectors $u = x+y$ and $v = x-y$. We apply Theorem \ref{thm:JL} to $u$ and $v$. By a union bound, except with probability $2\tau$ we have: $|||A(x+y)||_2^2 - ||x+y||_2^2| \leq \varsigma||x+y||_2^2$ and $|||A(x-y)||_2^2 - ||x-y||_2^2| \leq \varsigma||x-y||_2^2$. Therefore:
\begin{eqnarray*}
\langle (Ax),(Ay)\rangle  &=& \frac{1}{4}\left(\langle A(x+y), A(x+y)\rangle  - \langle A(x-y), A(x-y)\rangle \right) \\
&=& \frac{1}{4}\left(||A(x+y)||_2^2 + ||A(x-y)||_2^2 \right)\\
&\leq& \frac{1}{4}\left((1+\varsigma)||x+y||_2^2 - (1-\varsigma)||x-y||_2^2\right) \\
&=& \langle x, y\rangle  + \frac{\varsigma}{2}\left(||x||_2^2 + ||y||_2^2\right)
\end{eqnarray*}
An identical calculation shows that $\langle(Ax),(Ay)\rangle \geq \langle x, y\rangle  - \frac{\varsigma}{2}\left(||x||_2^2 + ||y||_2^2\right)$, which completes the proof.
\end{proof}

Instead of outputting synthetic data, our algorithm will output a datastructure based on a collection of random projections.
\begin{definition}[Projected Halfspace Data Structure]
A $T$ dimensional projected halfspace data structure of size $m$, $D_H = \{\{A_i\}, U, \{v_{i,j}\}\}$ consists of three parts:
\begin{enumerate}
\item $m$ independently selected random projection matrices $A_1,\ldots,A_m \in \mathbb{R}^{T\times d}$ mapping vectors from $\mathbb{R}^d$ to vectors in $\mathbb{R}^T$.
\item A collection of $T$-dimensional unit vectors $U \subset \mathbb{R}^T$
\item For each $i \in [m]$ and $j \in U$, a real number $v_{i,j} \in \mathbb{R}$.
\end{enumerate}
A projected halfspace data structure $D_H$ can be used to evaluate a halfspace query $f_y$ as follows. We write $f_y(D_H)$ to denote this evaluation:
\begin{enumerate}
\item For $i \in [m]$, compute the projection $\hat{y}_i \in \mathbb{R}^T$ as: $\hat{y}_i = A_i\cdot y$.
\item For each $i \in [m]$ compute $u_{i,j(i)} = \argmin_{u_j \in U} ||\hat{y}_i-u_j||_2$
\item Output $f_y(D_H) = \frac{1}{m}\sum_{i=1}^mv_{i,j(i)}$
\end{enumerate}
\end{definition}
What the projected halfspace data structure does is maintain $m$ projections into a low dimensional space, as well as a collection of `canonical' halfspaces $U$ in $T$ dimensions. The canonical halfspaces will be selected to form a net such that for every $\hat{y} \in \mathbb{R}^T$ with $||y||_2 = 1$, there is some $u \in U_i$ such that $||\hat{y}-u||_2 \leq \gamma/4$. The size of $U$ will be exponential in $T$, but we will choose $T$ to be only a constant so that maintaining such a set is feasible. Each $v_{i,j}$ will represent the approximate answer to the query $f_{u_j}$ on a projection of the private database by $A_i$. The Johnson-Lindenstrauss lemma will guarantee that not many points with margin $\gamma/2$ are shifted across the target halfspace by any particular projection, and the average of the approximate answers across all $m$ projections will with high probability be accurate for every halfspace.

First we bound the size of the needed net $U$ for halfspaces.
\begin{definition}
A $\gamma$-net for unit vectors in $\mathbb{R}^T$ is a set of points $U \subset \mathbb{R}^T$ such that for all $x \in \mathbb{R}^T$ with $||x||_2 = 1$:
$$\min_{y \in \mathbb{R}^T} ||x - y||_2 \leq \gamma$$
\end{definition}
\begin{claim}
There is a $\gamma$-net for unit vectors in $\mathbb{R}^T$ of size $|U| \leq \left(\frac{\sqrt{T}}{\gamma}\right)^T$.
\end{claim}
\begin{proof}
Consider the set of $T$ dimensional vectors discretized to the nearest multiple of $\gamma/\sqrt{T}$ in each coordinate. There are $\left(\frac{\sqrt{T}}{\gamma}\right)^T$ such vectors. For any unit $x \in \mathbb{R}^T$, let $y = \argmin_{y \in \mathbb{R}^T} ||x - y||_2$. We have: $||x-y||_2 \leq \sqrt{\sum_{i=1}^T(\gamma/\sqrt{T})^2} = \gamma$.
\end{proof}

We can now present our algorithm.
\begin{algorithm}
\textbf{ReleaseHalfspaces}($D, d, \gamma, \alpha, \eps$)
\begin{algorithmic}
\STATE \textbf{Let}:
 $$\varsigma \leftarrow \frac{\gamma}{4} \ \ \ \tau \leftarrow \frac{\alpha}{8} \ \ \ T \leftarrow \lceil 20\varsigma^{-2}\log(1/\tau) \rceil \ \ \ m \leftarrow \frac{16}{\alpha^2}d\left(\log(4\sqrt{d}/\gamma)+\log(1/\beta)\right)$$
\STATE \textbf{Let} $A_i \in \{-1/\sqrt{T},1/\sqrt{T}\}^{T\times d}$ be a uniformly random matrix for each $i \in [m]$.
\STATE \textbf{Let} $U$ be a $\gamma/4$-net for unit vectors in $\mathbb{R}^T$.
\FOR{$i = 1$ to $m$}
    \STATE \textbf{Let} $\hat{D}_i \subset \mathbb{R}^T$ be $\hat{D}_i = \{A_i x : x \in D\}$.
    \FOR{$x_j \in U$}
        \STATE \textbf{Let} $p_{i,j} \leftarrow \textrm{Lap}\left(\frac{m|U|}{\eps n}\right)$,  $v_{i,j} \leftarrow f_{x_j}(\hat{D}_i) + p_{i,j}$
    \ENDFOR
\ENDFOR
\STATE \textbf{Release} $D_H = (\{A_i\},U,\{v_{i,j}\})$.
\end{algorithmic}
\end{algorithm}

\begin{theorem}
ReleaseHalfspaces preserves $\epsilon$-differential privacy.
\end{theorem}
\begin{proof}
Privacy follows from the fact that the composition of $k$ $\epsilon$-differentially private mechanisms is $k\epsilon$-differentially private The algorithm makes $m|U|$ calls to the Laplace mechanism, and each call preserves $\epsilon/(m|U|)$-differential privacy (since each query has sensitivity $1/n$).
\end{proof}

\begin{theorem}
For any database $D$ with:
$$n \geq \frac{m(8\sqrt{T}/\gamma)^T}{\epsilon}\log\left(\frac{2m(\sqrt{T}/\gamma)^T}{\beta}\right)$$
Then except with probability at most $\beta$, $D_H = \textrm{ReleaseHalfSpaces}(D, d, \gamma, \alpha, \eps)$ is such that for each unit vector $y \in \mathbb{R}^d$ with margin $\gamma$ with respect to $D$: $|f_y(D)-f_y(D_H)| \leq \alpha$.
The running time of the algorithm and the bound on the size of $D$ are both polynomial for $\gamma, \alpha \in \Omega(1)$.
\end{theorem}
\begin{proof}
The analysis follows from the Johnson-Lindenstrauss lemma and a Chernoff bound.  Let $U^d$ be a $\gamma/4$-net for unit vectors in $\mathbb{R}^d$. Fix any $y \in \mathbb{R}^d$ such that $f_y$ has margin $\gamma$ with respect to $D$, and let $y' = \argmin_{y' \in U^d}||y-y'||_2$. Note that $f_{y'}$ has margin at least $\frac{3}{4}\gamma$ with respect to $D$ and that $f_y(D) = f_{y'}(D)$. Consider the quantity $|f_{y'}(D) - f_{y'}(D_H)|$, where $D_H$ is the projected halfspace data-structure output by ReleaseHalfspaces. By Corollary \ref{cor:JL}, for each $i \in [m]$ and each $x \in D$:
$$\Pr_A[|\langle(A_ix),(A_i{y'})\rangle - \langle x, y'\rangle | \geq \gamma/4] \leq \alpha/4$$
By linearity of expectation, the expected number of points in $D$ moved by more than $\gamma/4$ with respect to $y'$ in any projection is at most $\alpha n/4$:
$$E\left[|\{x \in D : f_{y'}(x) = f_{A_iy'}(A_ix) \wedge |\langle A_ix, A_iy' \rangle| \geq \frac{1}{2}\gamma\}|\right] \geq n\left(1-\frac{\alpha}{4}\right)$$
Let $u_{i,y'} = \argmin_{u \in U} ||u-A_iy'||_2$. Because $||u_{i,y'}-A_iy'||_2 \leq \gamma/4$, we have  for any $x \in \hat{D}_i$:
$$|\langle A_iy',x \rangle |= |\langle u_i, x \rangle + \langle A_iy'-u_{i,y'},x \rangle| \leq |\langle u_{i,y'}, x \rangle| + ||A_iy'-u_{i,y'}||_2||x||_2 \leq \langle u_i, x \rangle + \gamma/4$$
Thus:
$$E\left[|\{x \in D : f_{y'}(x) = f_{u_{i,y'}}(A_ix) \wedge |\langle A_ix, u_{i,y'} \rangle| \geq \frac{1}{4}\gamma\}|\right] \geq n\left(1-\frac{\alpha}{4}\right)$$
In other words, $f_{y'}(D) - \alpha/4 \leq E[f_{u_{i,y'}}(\hat{D}_i)] \leq f_{y'}(D) + \alpha/4$. Moreover, for each $i$, $f_{u_{i,y'}}(\hat{D}_i)$ is an independent random variable taking values in the bounded range $[0,1]$, and so we will be able to apply a Chernoff bound. For each $y'$:
$$\Pr[|\frac{1}{m}\sum_{i=1}^mf_{u_{i,y'}}(\hat{D}_i) - E[f_{u_{y'}}(\hat{D})]| \geq \frac{\alpha}{2}] \leq 2\exp\left(-\frac{m\alpha^2}{2}\right)$$
Taking a union bound over all $(4\sqrt{d}/\gamma)^d$ vectors $y' \in U^d$ and plugging in our chosen value of $m$, and recalling our bound on $E[f_{u_{y'}}(\hat{D})]$ we find that:
$$\Pr[\max_{y' \in U^d}|\frac{1}{m}\sum_{i=1}^mf_{u_{i,y'}}(\hat{D}_i) -  f_{y'}(D)| \geq \frac{3\alpha}{4}] \leq \frac{\beta}{3}$$
Also note that the algorithm makes $m|U|$ draws from the distribution $\textrm{Lap}\left(\frac{m|U|}{\eps n}\right)$ during its run, assigning these draws to values $p_{i,j}$. Except with probability at most $\beta/3$, we have for all $i, j$:
$$|p_{i,j}| \leq \frac{m|U|}{\epsilon n}\log\left(\frac{2m|U|}{\beta}\right) \leq 1$$
Therefore, conditioning on $|p_{i,j}| \leq 1$ for all $i,j$ and  applying another Chernoff bound, we find that for any sequence of indices $j(i)$:
$$\Pr[|\frac{1}{m}\sum_{i=1}^mp_{i,j(i)}| \geq \alpha/4] \leq 2\exp\left(-\frac{m\alpha^2}{8}\right)$$
Again taking a union bound and plugging in our value of $m$, we find that:
$$\Pr[\max_{j(1),\ldots,j(m)}|\frac{1}{m}\sum_{i=1}^mp_{i,j(i)}| \geq \alpha/4] \leq \frac{\beta}{3}$$
Finally, conditioning on these three events (which together occur except with probability $\beta$), we have for any $y'$:
$$f_{y'}(D_H) = \frac{1}{m}\sum_{i=1}^mv_{i,j(i)} = \frac{1}{m}\left(\sum_{i=1}^mf_{u_{i,y'}}(\hat{D}_i) + \sum_{i=1}^m p_{i,j(i)}\right) \leq f_{y'}(D)+ \alpha$$
which completes the proof.
\end{proof}

\section{Distributional Privacy}\label{sec:distributional}
In this section we give an alternative privacy definition, motivated by learning theory, and show that it is a strengthening of differential privacy.

\begin{definition}[$S$-neighbors]
For any subset of the universe $S \subseteq X$ of size $|S| \geq n$, we say that two databases are $S$-neighbors if they were both drawn at random \emph{without replacement} from $S$.
\end{definition}

\begin{definition}[Distributional Privacy]
We say that a mechanism $A:X^n\rightarrow R$ satisfies
\emph{$(\epsilon,\beta)$-distributional privacy} if for any $S \subseteq X$, and for any pair of databases $D_1,D_2 \subseteq X$ of size $n$ which are $S$-neighbors, with probability $1-\beta$ over the draw of the databases we have for all events $E \subseteq R$:
$$\Pr[A(D_1) \in E] \leq e^{\epsilon}\Pr[A(D_2)\in E].$$
\end{definition}

{For example, suppose that a collection of hospitals in a region
each treats a random sample of patients with disease $X$.
A hospital can release information with a guarantee of distributional privacy, which is informative about patients in the region, without revealing which hospital the data
came from. Actually, our main motivation is that this definition
is particularly natural from the perspective of learning theory:
given a sample of points drawn from some distribution $\D$, one
would like to reveal no more information about the sample than is
inherent in $\D$ itself.}

We will typically think of $\beta$ as being exponentially small,
whereas $\epsilon$ must be $\Omega(1/n)$ for $A$ to be useful.

\subsection{Relationship Between Definitions}
It is not a priori clear whether either differential privacy or
distributional privacy is a stronger notion than the other, or if the
two are equivalent, or distinct. On the one hand, differential privacy
only provides a guarantee when $D_1$ and $D_2$ differ in a single
element,\footnote{We get $t\epsilon$-differential
privacy for $D_1$ and $D_2$ that differ in $t$ elements.} whereas
distributional privacy can provide a guarantee for two databases $D_1$
and $D_2$ that differ in all of their elements. On the other hand,
distributional privacy makes the strong assumption that the elements in
$D_1$ and $D_2$ are drawn from some distribution $\mathcal{D}$, and allows for
privacy violations with some exponentially small probability $\beta$
(necessarily: with some small probability, two databases drawn from the
same distribution might nevertheless possess very different statistical properties). However,
as we show, distributional privacy is a strictly stronger guarantee than
differential privacy.

\begin{theorem}
If $A$ satisfies $(\epsilon,\beta)$-distributional privacy for any
$\beta = o(1/n^2)$, then $A$ satisfies $\epsilon$-differential privacy.
\end{theorem}
\begin{proof}
Consider any database $D_1$ drawn from domain $X$, and any neighboring
database $D_2$ that differs from $D_1$ in only a single element $x \in
X$. Let $S = D_1 \cup \{x\}$ be a set of size $|S| = n+1$. Consider two $S$-neighbors $D_1'$ and $D_2$', then
with probability $2/n^2$ we have $\{D_1',D_2'\} = \{D_1,D_2\}$, and so
if $\beta = o(1/n^2)$, we have with certainty that for all events $E$:
$$\Pr[A(D_1')\in E] \leq e^\epsilon\Pr[A(D_2') \in E].$$
Since this holds for all pairs of neighboring databases, $A$ satisfies $\epsilon$-differential privacy.
\end{proof}

\begin{definition}
Define the {\em{mirrored mod $m$ function}} as follows:
$$F_m(x) = \left\{
             \begin{array}{ll}
               x \mod m, & \hbox{if $x \mod 2m < m$;} \\
               -x-1 \mod m, & \hbox{otherwise.}
             \end{array}
           \right.$$
\end{definition}

For a database $D \subset \{0,1\}^n$, define the query
$$Q_m(D) = F_m(\sum_{i = 1}^{n}D[i]).$$
Note that the global sensitivity of any query $Q_m$ satisfies $GS_{Q_m}
\leq 1$. Therefore, the mechanism $A$ that answers queries $Q_n$ by
$A(D,Q_m) = Q_m(D) + Z$ where $Z$ is drawn from
$\textrm{Lap}(1/\epsilon)$ satisfies
$\epsilon$-differential privacy.

\begin{theorem}
There exist mechanisms $A$ that satisfy $\epsilon$-differential privacy, but
do not satisfy $(\epsilon,\beta)$-distributional privacy for any $\epsilon < 1,$
$\beta = o(1)$ (that is, for any meaningful values of
$\epsilon,\beta$).
\end{theorem}
\begin{proof}
Consider databases with elements drawn from $X = \{0,1\}^n$ and
the query $Q_{2/\epsilon}$. As observed above, a mechanism $A$ such that
$A(D,Q_i) = Q_i(D) + Z$ for $Z \sim \textrm{Lap}(1/\epsilon )$ satisfies
$\epsilon$-differential privacy for any $i$. Note however that with
constant probability, two databases $D_1,D_2$ drawn from $S = X$ have
$|Q_{2/\epsilon}(D_1)-Q_{2/\epsilon}(D_2)| \geq 1/\epsilon $. Therefore,
for any output $x$, we have that with constant probability over draws of two $S$ neighbors $D_1$ and $D_2$:
\begin{align*}\frac{\Pr[A(D_1,Q_{2/\epsilon}) = x]}{\Pr[A(D_2,Q_{2/\epsilon}) = x]}&=
e^{-\epsilon|Q_{2/\epsilon}(D_1)-Q_{2/\epsilon}(D_2)|} \\
&=
e^{-\epsilon (\frac{1}{\epsilon })} \\&= \frac{1}{e}~.\qed
\end{align*}
Therefore the mechanism does not satisfy $(1,o(1))$-distributional privacy.
\end{proof}

Although there are simpler functions for which preserving distributional
privacy requires more added noise than preserving differential privacy,
the mirrored-mod function above is an example of a function for which it
is possible to preserve differential privacy usefully, but yet
impossible to reveal any useful information while preserving distributional
privacy.

We note that in order for distributional privacy to imply differential
privacy, it is important that in the definition of distributional
privacy, database elements are drawn from some distribution $\mathcal{D}$
\textit{without replacement}. Otherwise, for any non-trivial
distribution, there is some database $D_*$ that is drawn with
probability at most $1/2^n$, and we may modify any
distributional-privacy preserving mechanism $A$ such that for every
query $Q$, $A(D_*,Q) = D_*$, and for any $D_i \ne D_*$, $A(D_i,Q)$
behaves as before. Since this new behavior occurs with probability $\leq
\beta$ over draws from $D$ for $\beta = O(1/2^n)$, $A$ still preserves
distributional privacy, but no longer preserves differential privacy
(which requires that the privacy guarantee hold for \textit{every} pair
of neighboring databases).

\section{Conclusions and Open Problems}
In this paper we have shown a very general information theoretic result: that small nets are sufficient to certify the existence of accurate, differentially private mechanisms for a class of queries. For counting queries, this allows algorithms which can accurately answer queries from a class $C$ given a database that is only \emph{logarithmic} in the size of $C$, or linear its VC-dimension. We then also gave an efficient algorithm for releasing the class of interval queries on a discrete interval, and for releasing large-margin halfspace queries in the unit sphere.

The main question left open by our work is the design of algorithms which achieve utility guarantees comparable to our net mechanism, but have running time only polynomial in $n$, the size of the input database. This question is extremely interesting even for very specific classes of queries. Is there such a mechanism for the class of conjunctions? For the class of parity queries?

\section{Acknowledgments} We thank David Abraham, Cynthia Dwork, Shiva
Kasiviswanathan, Adam
Meyerson, Ryan O'Donnell, Sofya Raskhodnikova, Amit Sahai, and Adam Smith for many useful
discussions.

\bibliographystyle{alpha}
\bibliography{journalversion}

\end{document}